\documentclass[natbib,final]{svjour3}                     
\smartqed  
\usepackage{graphicx,epsfig,fancyhdr,rotating}
\usepackage{txfonts}
\usepackage{mathptmx}      
\newcommand{\degree}{\ensuremath{^\circ}}
\newcommand{\arcsec}{^{''}}

\journalname{Space Science Review}
\begin{document}
\title{Propagating MHD waves in coronal holes
}

\titlerunning{Propagating MHD Waves in Coronal holes}        

\author{D. Banerjee   \and G. R. Gupta      \and
        L. Teriaca 
}


\institute{D. Banerjee \at
              Indian Institute of Astrophysics, Koramangala, Bangalore 560034, India \\
              Tel.: +91 80 25530672\\
              Fax: +91 80 25534043\\
              \email{dipu@iiap.res.in}             \\
\\
G. R. Gupta \at
	  Indian Institute of Astrophysics, Koramangala, Bangalore, 560034, India \\
          Joint Astronomy Programme, Indian Institute of Science, Bangalore, 560012, India\\
          \email{girjesh@iiap.res.in}             \\
\\
	  L. Teriaca \at
         Max-Planck-Institut f\"{u}r Sonnensystemforschung (MPS), 37191,  Katlenburg-Lindau, Germany\\
         \email{teriaca@linmpi.mpg.de}             \\ 
}

\date{Received: \today / Accepted: date}

\maketitle
\begin{abstract}
Coronal holes are the coolest and darkest regions of the solar atmosphere, as observed both on the solar disk and above the solar limb. Coronal holes are associated with rapidly expanding open magnetic fields and the acceleration of the high-speed solar wind. During the years of the solar minima, coronal holes are generally confined
to the Sun's polar regions, while at solar maxima they can also be found
at lower latitudes. Waves, observed via remote sensing and detected in-situ in the wind streams, are most likely responsible for the wind and several theoretical models describe the role of MHD waves in 
the acceleration of the fast solar wind. This paper reviews the observational evidences of detection of propagating waves in these regions. The characteristics of the waves, like periodicities, amplitude, speed provide input parameters and also act as  constraints on theoretical models of coronal heating and solar wind acceleration. 

\end{abstract}

\section{Introduction}
\label{intro}

Coronal holes are regions of cool and low density plasma that appear
`dark' when seen in lines formed at coronal temperatures \citep{1972ApJ...176..511M}. 
On the other hand, they are almost indistinguishable from
their surroundings in photospheric and low chromospheric temperatures and there is no significant radiance
contrast between the holes and the surrounding region until the temperature exceeds $10^{5}$ K.
 During the years of the solar minima, coronal holes are generally confined
to the Sun's polar regions, while at solar maxima they can also be found
at lower latitudes. Coronal holes are observed both on disk and above the solar limb (off-limb). They are associated with rapidly expanding
open magnetic fields and with the acceleration of the high-speed solar wind. For a recent review on coronal holes see \citet{2009LRSP....6....3C}. During the solar minimum, Ulysses observations clearly show that
the solar wind exhibits two modes of outflow: the fast wind, associated with 
polar coronal holes, with outflow speeds of $\approx 800$~km~s$^{-1}$ and the 
slow  wind with outflow speeds of $\approx400$~km~s$^{-1}$\ associated with 
equatorial regions \citep{1997GeoRL..24.2885W,2000JGR...10510419M}. However, during solar maxima, low latitude coronal
 holes also show faster than average solar wind speed upto 
 $\approx 600$~km~s$^{-1}$ \citep{2003JGRA..108.1144Z}. 

  There are several theoretical models which describe the role of MHD waves in 
the acceleration of the fast solar wind in coronal holes. Some of these models were investigated
 using 1.5D MHD equations \citep{1996A&A...308..299B,1996ApJ...465..451L,1996JGR...10115615S,2000A&A...353..741N,2001ApJ...554.1151S,2002ApJ...571L.187L,2004MNRAS.349.1227S,2005ApJS..156..265C},
2.5D MHD equations \citep{1995JGR...10023413O,1997ApJ...476L..51O}, 2.5D multifluid MHD equations
\citep{2001ApJ...553..935O,2004JGRA..10907102O}. The common thread in the above models of the fast
solar wind acceleration is the required presence of nonlinear MHD waves, or shock wave trains.
Some of these wave driven wind models have been reviewed by \citet{2005SSRv..120...67O}.

 The detection of these waves in the outer
solar atmosphere is made possible by analyzing the effects these waves have on the plasma. The
presence or signature of compressional waves may be seen in the form of variations or oscillations
in radiance, due to change in plasma density, and also in the
line-of-sight (LOS) velocities, due to plasma motions (when they have a significant component directed towards the observer). On the other hand, transverse waves give rise to only LOS effects when they propagate substantially perpendicular to the observer. Moreover the latter give no radiance signature in the theoretical limit of uncompressible Alfv\'en waves.
Temporally and spatially resolved motions result in shifts of the observed profiles while unresolved motions result in broadening of the spectral lines.
These effects can be measured from the spectroscopic studies of spectral lines. 

Solar and Heliospheric
 Observatory \citep[SoHO,][]{1995SoPh..162....1D} and Hinode \citep{2007SoPh..243....3K} data shed lights onto the dynamical events such as short time scale
variability or oscillations in the coronal hole atmosphere observed at Vacuum UltraViolet (VUV: 100~$\AA$\ 
 to 1600 \AA) wavelengths.
These periodic oscillations generally carry information from the emitting regions which allows us to
diagnose the frozen-in magnetic fields as well as the plasma contained in the different magnetic 
structures (e.g., plumes, magnetic networks). The wavelengths of these waves are often comparable
 to the characteristic sizes of these coronal structures and measured time scales are in the range
 of seconds to few minutes. Table~\ref{tab:list} gives the overview of the periodicities and 
propagation speeds of propagating MHD waves detected in these coronal hole structures. In this review we focus on the observational evidences of waves at different locations in coronal holes (on disk and off-limb) which can put constraints on theoretical models of coronal heating and solar wind acceleration.

\begin{table}
\caption{Overview of the periodicities and 
propagation speeds of propagating MHD waves detected in coronal hole structures by remote sensing.}
\label{tab:list}       
\begin{tabular}{lllll}
\hline\noalign{\smallskip}
Authors & Regions & Periods & Speed & Instrument \\
	&      &   (s)	&  (km~s$^{-1}$) &  \\
\noalign{\smallskip}\hline\noalign{\smallskip}
\citep{1983SoPh...89...77W} & Off-limb & --   &100-200& Skylab\\
\citep{1997ApJ...491L.111O} & Off-limb & 360 & -- & UVCS \\
\citep{1998ApJ...501L.217D} & Plumes   & 600-900 & 75-150 & EIT \\
\citep{2000ApJ...529..592O} & Off-limb & 400-625 & 160-260 & UVCS \\
\citep{2000SoPh..196...63B} & Plume & 600-1200 & -- & CDS \\
{\citep{2001A&A...377..691B}} & Inter-plume & 1200-1800 & -- & CDS \\ 
{\citep{2001A&A...380L..39B}} & On-disk & 600-1200 & -- & CDS \\
{\citep{2002A&A...393..649M}} & On-disk & 100-900 and 1500 & -- & CDS \\
{\citep{2005A&A...442.1087P}} & Off-limb & 600-5400 and 10200 & -- & SUMER \\
{\citep{2006A&A...452.1059O}} & Off-limb & 300-1000 & 150-170 & CDS \\
{\citep{2007A&A...463..713O}} & On-disk & 300-1000 & 50-70 & CDS \\
{\citep{2008ApJ...677L.137B}} & Off-limb & 10000-350000  &--& UVCS \\
{\citep{2008A&A...488..331T}} & On-disk & 300-600 & 30-60 & TRACE \\
{\citep{2009A&A...493..251G}} & On-disk & 300-1500 & 4-15 & SUMER \\
{\citep{2009A&A...499L..29B}} & Plume & 600-1800 & 75-125 & EIS-SUMER \\
\citep{2010ApJ...718...11G}   & Off-limb & 600-1200 & 25-330 & EIS-SUMER \\

\noalign{\smallskip}\hline
\end{tabular}
\end{table}

\begin{figure}[htb]
\centering
\includegraphics[angle=90,width=10cm]{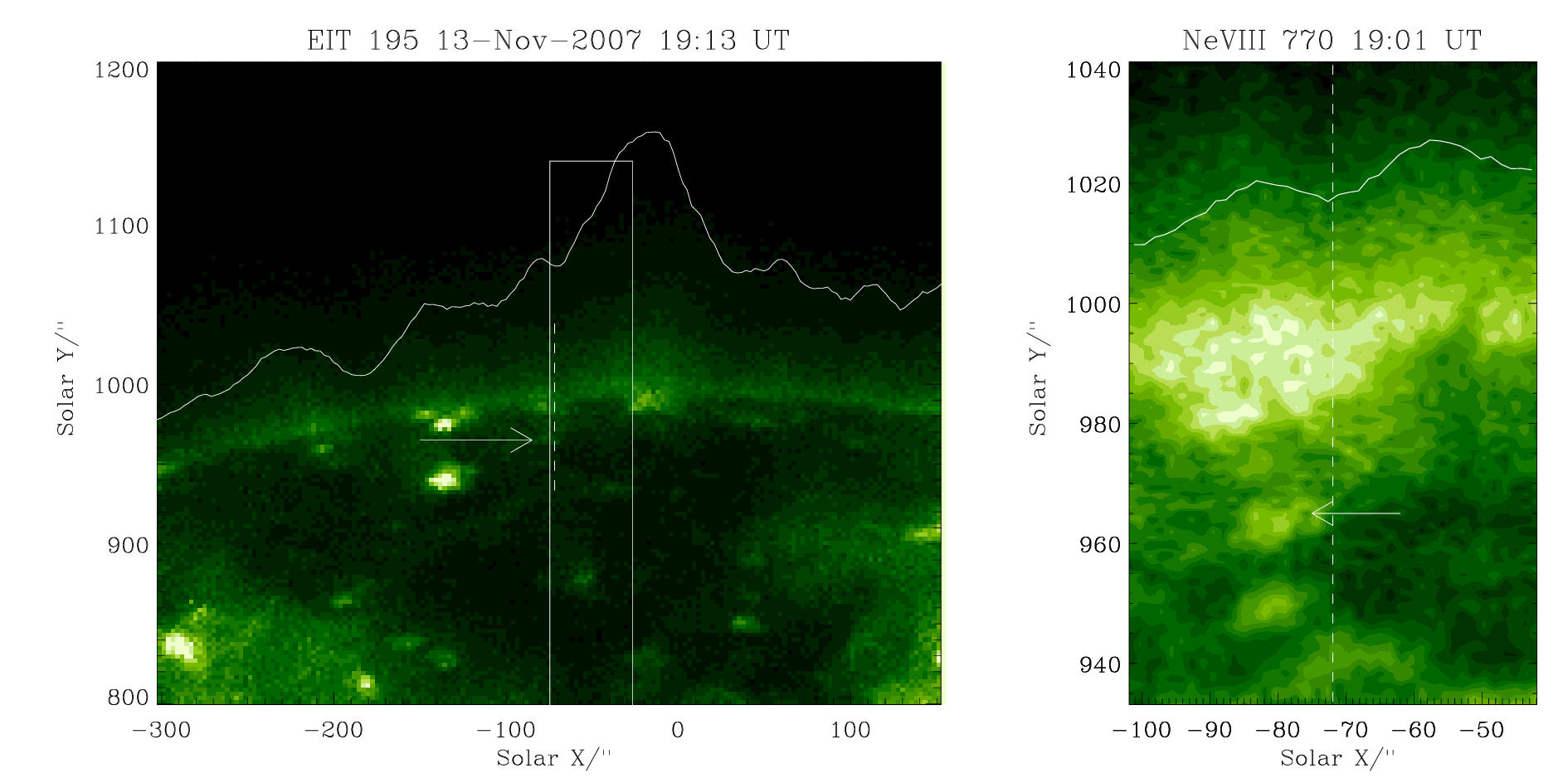}
\caption{Left: plume and inter-plume region visible on the EIT image taken on $13^{th}$
 November in the Fe~{\sc xii}~195~\AA\ passband. The radiance variation along the solar-X is
 over-plotted as a white line in an arbitrary unit at fixed solar-Y$\approx1000\arcsec$. 
This variation along the solar-X allows to identify the location of plumes and inter-plumes.
The rectangular box marks the location of the EIS slot while the dashed line gives the
 location of the SUMER slit. Right: The SUMER context raster taken in Ne~{\sc viii} 770~\AA\
 spectral line. The continuous line gives the radiance variation along the 
solar-X in an arbitrary unit. In both the panels, the arrow indicates the location of the 
bright region from where waves are presumably originating. Adapted from \citet{2010ApJ...718...11G}.}
\label{fig:map}
\end{figure}

\section{Observational evidence of waves in Coronal Holes}
Evidence for Alfv\'{e}nic fluctuations, and MHD waves is obtained by \textit{in situ} and
remote sensing observations throughout the heliosphere. Here we will describe few representative observations
from low to high heliocentric distances. In the photosphere, the high resolution time series observations of G-band bright points in 
intergranular lanes show the presence of transverse motions. The mean transverse velocities inferred
 from these observations can be as large as 1.4~km~s$^{-1}$, but only some fraction of that will be
 likely to propagate upwards on open field lines \citep[e.g.,][]{1994A&A...283..232M,2003ApJ...587..458N}.
\citet{1998A&A...339..208B,2009A&A...501L..15B} have observed nonthermal broadening of spectral lines from 
various ions using the SUMER/SoHO \citep{1995SoPh..162..189W} and EIS/Hinode \citep{2007SoPh..243...19C}
 in off-limb coronal holes. The authors assumed that
spectral lines are formed in a plasma with an electron temperature $T_{\rm e}$ 
equal to the peak temperature of the equilibrium ionization balance for each
ion. In this context, broader lines imply an additional
 unresolved velocity due to wave motions along the line of sight (for further details see section 3.2).  
 At larger heights, measurements of the visible polarization brightness (pB) with the Ultra-Violet Coronagraph
Spectrometer \citep[UVCS/SoHO,][]{1995SoPh..162..313K}
 White Light Channel were also used to infer the presence of density oscillations \citep{1997ApJ...491L.111O}.
Interplanetary scintillation (IPS) observations of radio signals passing through the corona allow
 some properties of plasma irregularities to be determined.
One way of detecting random fluctuations in the
bulk solar wind is by measuring departures from a
frozen-in diffraction pattern measured by different
sets of radio receivers. \citet{1981A&A...103..415A} made an early attempt
to separate the bulk solar wind flow speed from the
random wavelike velocity component within the 30~R/R$_{\odot}$.
Radio IPS measurements are sensitive to density
fluctuations over a wide range of spatial scales.
\citet{2002ApJ...576..997S} presented integrated values of
$\delta\rho/\rho_{o}$ from VLBI measurements, and compared
them to predictions based on specific MHD modes.
The Helios 1 and 2 probes uniquely measured the
\textit{in situ} plasma properties between Mercury and the
Earth, and they measured MHD fluctuations spanning
a wide range of time scales \citep[e.g.,][]{1995SSRv...73....1T,1995ARA&A..33..283G}.
The \textit{in situ} density fluctuation spectra between 0.3
and 1 AU show a large intrinsic variability, with no
clear radial trend discernible \citep{1994JGR....9921481T}.
Ulysses spacecraft has detected a spectrum of outward propagating and inward propagating
waves in the millihertz frequency range with power laws of $f^{-1}$, at low frequency
($<10^{-4}$ Hz) and $f^{-5/3}$ for higher frequencies beyond 1 AU \citep{1995GeoRL..22.3393G},
as well as large amplitude ultra low frequency Alfv\'{e}n waves \citep{1995GeoRL..22.3397T}.

\section{Detection of propagating waves in polar plumes and inter-plumes}
\label{sec:xtmap}

The study of polar plumes, the unipolar high density structures in
coronal holes, provides clues to the understanding of solar wind
acceleration and coronal heating. Plumes subtend an angle of 2\degree\ relative to the Sun centre
 at low latitude and expand
super-radially with the coronal hole \citep{1968SoPh....3..321N,1997SoPh..175..393D}.
The region between these structures is termed as inter-plume. 
From VUV spectroscopy, plumes are known to be denser and cooler than the 
surrounding inter-plume regions \citep[e.g.,][]{2006A&A...455..697W}, while 
spectral lines are observed to be broader in inter-plumes 
\citep[i.e.,][]{2000SoPh..194...43B,2000ApJ...531L..79G,2003ApJ...588..566T}.
  \citet{1999AAS...19410801D}
produced images from LASCO/SoHO \citep{1995SoPh..162..357B}, which clearly show polar plumes
extending to altitudes of 25~R/R$_{\odot}$ or more, very close to
the outer edge of the C-3 field of view and above the likely
Alfv\'enic point of the wind flow. Figure~\ref{fig:map} shows plume and inter-plume region in
coronal hole where the radiance variation along the solar-X at 
solar-Y$\approx1000\arcsec$ is over-plotted as a white line in arbitrary units which allows to 
identify the location of plumes and inter-plumes in the region within field of view. There is some
indication that pressure balance structures which are a common feature observed \textit{in situ}
 in high latitude, fast solar wind near solar minimum are remnants of coronal polar
plumes \citep{2002GeoRL..29j..21Y}. Plumes are regions of plasma density enhancements.
Some of the studies concluded that plumes have lower outflow speeds than inter-plume regions 
\citep{1997AdSpR..20.2219N,2000ApJ...531L..79G,2000A&A...353..749W,2000A&A...359L...1P,2003ApJ...588..566T,2007ApJ...658..643R} 
and, hence, may not contribute significantly to the fast solar wind, whereas some other theoretical 
and observational studies find higher outflow speeds in plumes than in inter-plume
regions for at least some altitudes above the photosphere \citep{1999JGR...104.9947C,2003ApJ...589..623G,2005ApJ...635L.185G}. 
These contradictory reports led to the debate on whether plumes or inter-plumes are the 
preferred source regions for the acceleration of the fast solar wind. 
This topic is highly debated and still open for further confirmation. Polar plumes
may also act as Alfv\'{e}n and slow wave guides. Thus, the study of MHD wave activity in
polar plumes is an important and interesting branch of solar coronal physics.\\

\begin{figure}[htbp]
 \centering
{\includegraphics[width=8cm]{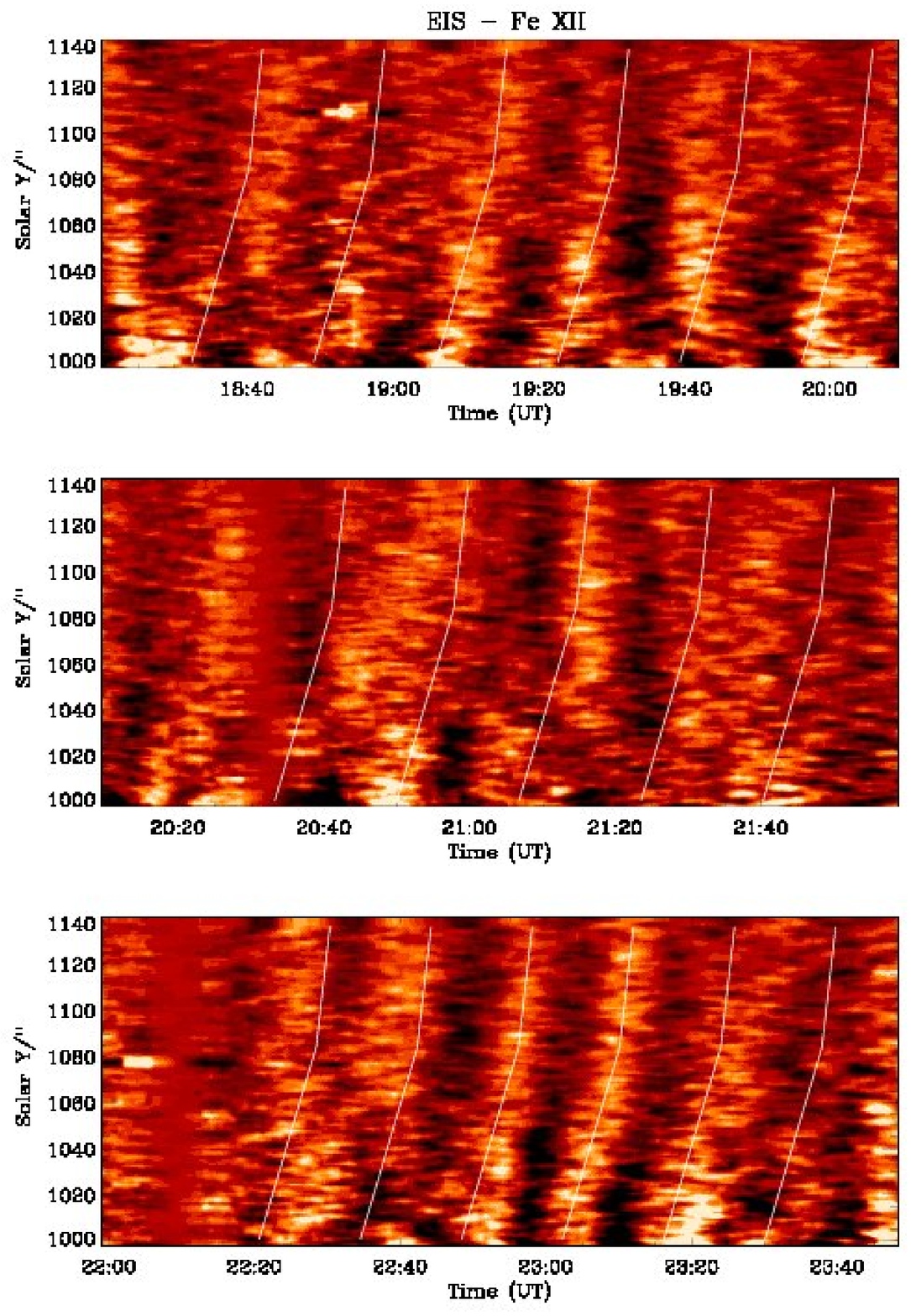}}\\
\caption{Enhanced x-t map of radiance variation along solar-Y at
 solar-X$\approx-72\arcsec$ as recorded by EIS in Fe~{\sc xii} 195~\AA\ on 
$13^{th}$ November 2007. The height range shown here covers the near off-limb and far off-limb region 
of the North polar coronal hole and corresponds to the inter-plume region. 
The slanted lines correspond to the disturbances propagating outward with increasing speed. 
In the near off-limb region the disturbance propagates with speed of ($130 \pm 14$)~km~s$^{-1}$, 
and accelerates to ($330 \pm 140$)~km~s$^{-1}$ in the far off-limb region. The period is in
 the range of $\approx$15 min to 18 min. From \citet{2010ApJ...718...11G}.
\label{fig:xt_fe12}}
\end{figure}

\begin{figure}[htbp]
 \centering
{\includegraphics[width=8cm]{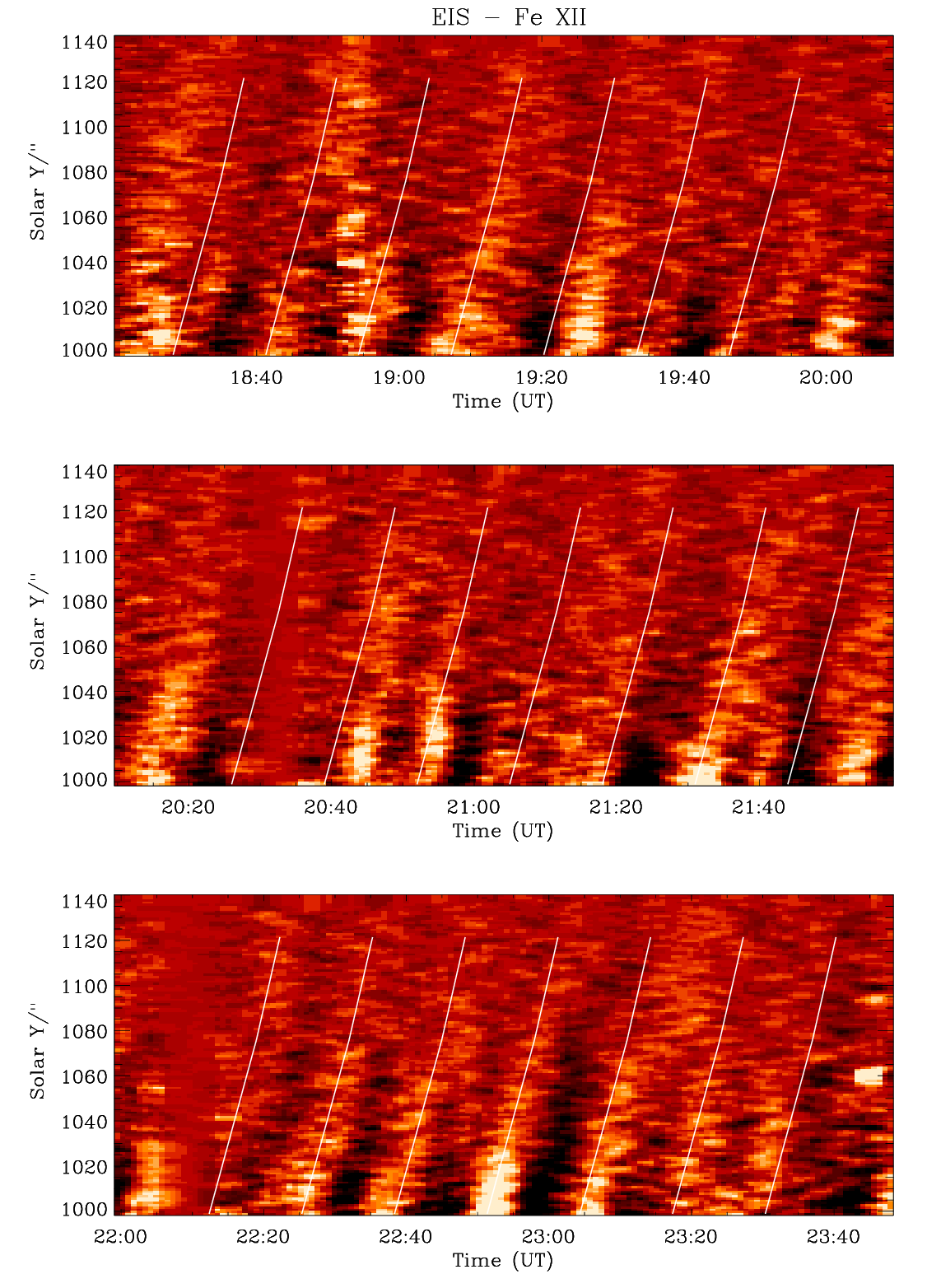}}\\
\caption{Enhanced x-t map of radiance variation along solar-Y at
 solar-X$\approx-39\arcsec$ as recorded by EIS in Fe~{\sc xii} 195~\AA\ on 
$13^{th}$ November 2007. The height range shown here covers the near off-limb and far off-limb region of 
the polar coronal hole and falls in the plume region. The slanted lines corresponds to the disturbances 
propagating outward with nearly constant speed. 
In the near off-limb region the disturbance propagates with speed of ($135 \pm 18$)~km~s$^{-1}$, and accelerates 
 to ($165 \pm 43$)~km~s$^{-1}$ in the far off-limb region. The period 
 is in the range of $\approx$15 min to 20 min. From \citet{2010ApJ...718...11G}.
\label{fig:xt_fe12p}}
\end{figure}

\subsection{Compressional waves}
The first observational indication of the presence of compressional perturbations
in polar plumes was reported by \citet{1983SoPh...89...77W} from Skylab observations. Statistically significant short period variations of
Mg~{\sc x} 625~\AA\ line radiance were detected, with propagating speed in the range of 100~km~s$^{-1}$ to 200~km~s$^{-1}$ and relative amplitudes  of about 10~\%.
Post SoHO launch, the first undoubted
detection of propagating slow MHD waves was made by UVCS/SoHO. Detection of slow waves in an
open magnetic structure high above the limb of coronal holes was
reported by \citet{1997ApJ...491L.111O,2000ApJ...529..592O}. \citet{1998ApJ...501L.217D},
 analysing Extreme-ultraviolet Imaging Telescope \citep[EIT/SoHO,][]{1995SoPh..162..291D}
data of polar plumes, detected similar compressive disturbances with relative
linear amplitudes of the order of 10~\% to 20~\% and periods of 10 min to 15
min. \citet{1999ApJ...514..441O,2000ApJ...533.1071O}
 identified the observed
compressive longitudinal disturbances as propagating slow MHD waves. \citet{2007SoPh..246....3B}
have summarized the main features of the observed oscillations within open magnetic structures. 
A number of studies using Coronal Diagnostic Spectrometer \citep[CDS/SoHO,][]{1995SoPh..162..233H}
 and SUMER/SoHO have reported
oscillations in plumes, interplumes and coronal holes in the polar
regions of the Sun \citep[e.g.,][]{2000SoPh..196...63B,2001A&A...380L..39B,2006A&A...452.1059O,2007A&A...463..713O}
. All of these
studies point to the presence of compressional waves, thought to be
slow magnetoacoustic waves as found by \citet{1998ApJ...501L.217D}.
The detected damping of slow propagating waves was attributed to
compressive viscosity.

\begin{figure*}[htb]
\centering
\includegraphics[width=12.5cm]{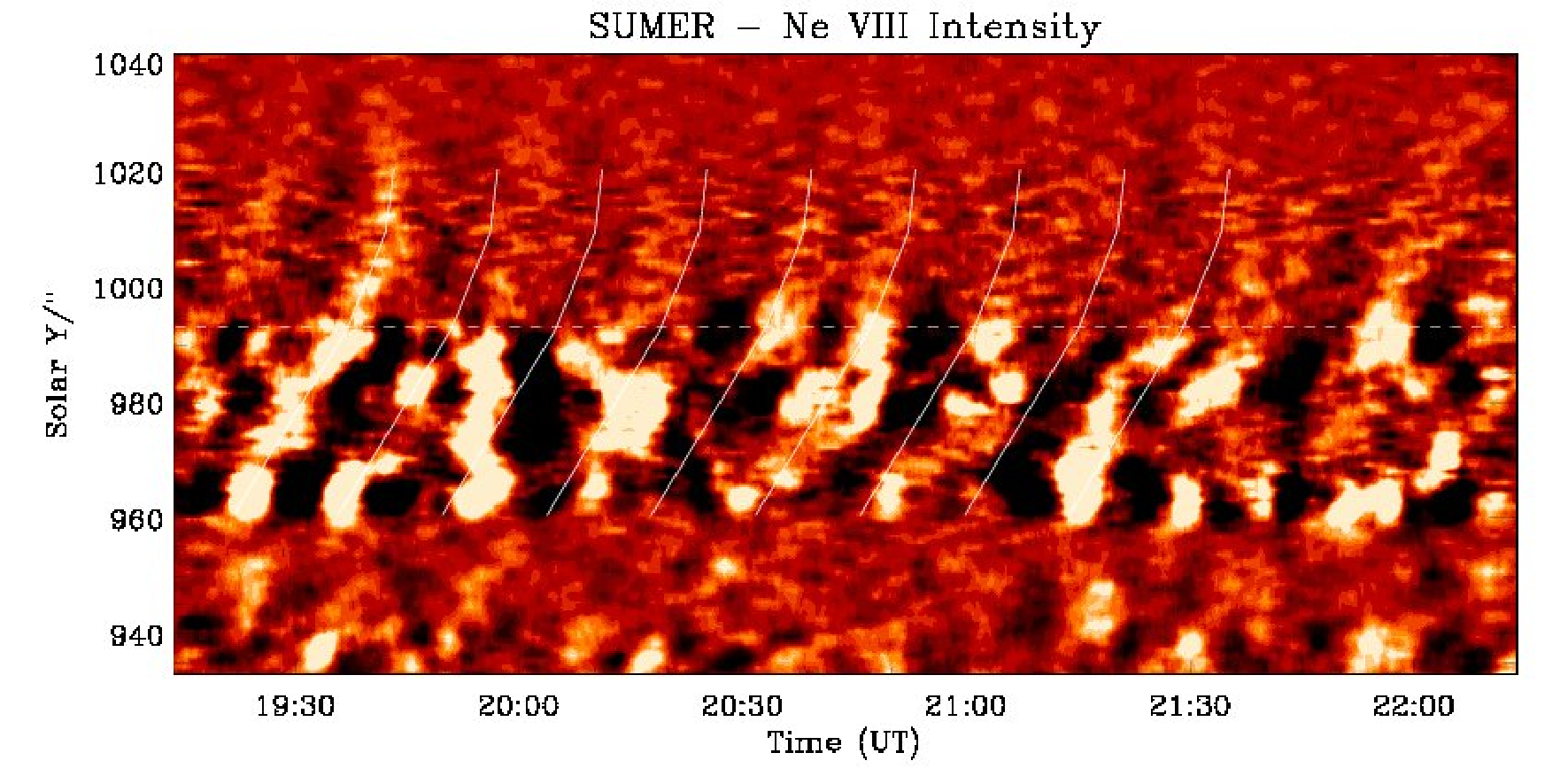}
\caption{Enhanced time-distance (x-t) map of radiance (along solar-Y) variation 
at solar X$\approx -72\arcsec$ as recorded by SUMER in the Ne~{\sc viii} 770~\AA\
spectral line on $13^{th}$ November 2007. Here the slit covers the on-disk, 
limb and off-limb region of the polar coronal hole and it is positioned in the 
inter-plume region. 
The slanted lines correspond to the disturbances propagating outward with 
increasing speed. The dashed horizontal line indicates the position of the limb 
brightening in Ne~{\sc viii} 770~\AA\.
In the on-disk region the disturbance propagates with a speed of $\approx25 \pm 1.3$ km s$^{-1}$, 
increasing to ($38 \pm 4.5 $)~km~s$^{-1}$ close to the limb and to about ($130 \pm 51$)~km~s$^{-1}$
 in the off-limb region. The period is in the range of
 $\approx$~14 min to 20 min. From \citet{2010ApJ...718...11G}.}
\label{fig:xt_ne8}
\end{figure*}

\begin{figure}[!th]
\centering
{\includegraphics[width=13cm, clip=true]{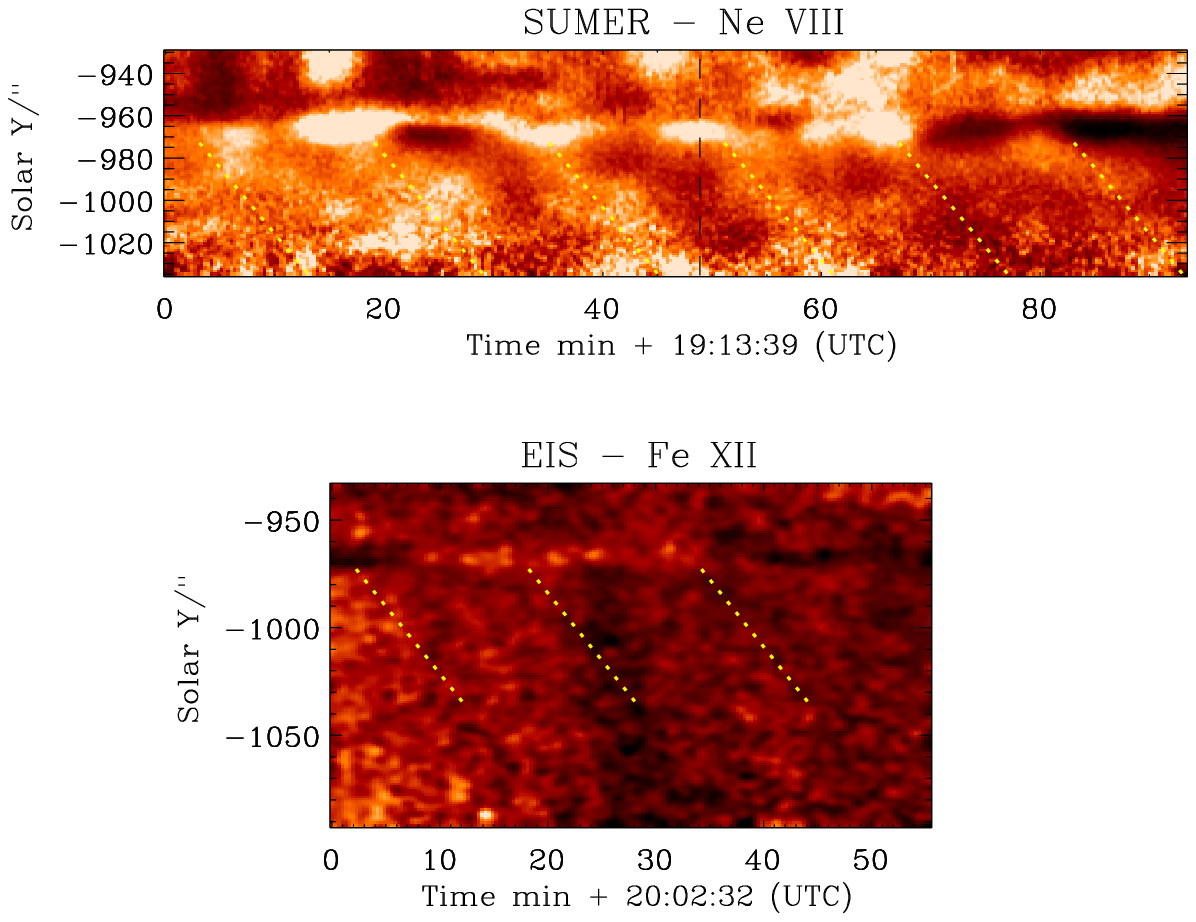}}\\
\caption{Enhanced maps of radiance variation along the slit 
(Solar Y direction) with time for Ne~{\sc viii} 770~\AA\ as recorded by SUMER 
(top panel) and Fe~{\sc xii} $195$~\AA\ as recorded by EIS (bottom panel) on 
$8^{th}$ April 2007 for the south polar coronal hole. The vertical black dashed line on the SUMER
enhanced radiance map depicts the starting point of the EIS time series (shown 
in the bottom panel). The slanted dotted yellow lines correspond to
disturbances propagating with a speed of 75~km~s$^{-1}$ and a period of about
15 min, as determined from the SUMER data. From \citet{2009A&A...499L..29B}.}
\label{fig:XTapr}
\end{figure}

Recently, \citet{2010ApJ...718...11G} have studied the presence of propagating waves in both the plume
and inter-plume region simultaneously. For this study, the data from SUMER 
and EIS spectrometer has been obtained simultaneously in space and time in several
different spectral lines. They have obtained time-distance (x-t) radiance maps in the Fe~{\sc xii}
 195~\AA\ spectral line in both regions simultaneously. The x-t map in inter-plume region shows the
signature of acceleration while propagating, see Figure~\ref{fig:xt_fe12}, while that for plume
region do not show such acceleration, see Figure~\ref{fig:xt_fe12p}. The measured propagation speed in interplume region was
($130\pm 14$)~km~s$^{-1}$ just above the limb, which accelerates to ($330\pm140$)~km~s$^{-1}$\ around
 $160\arcsec$ above the limb. Whereas in plume region, the propagation 
speed was in the range of ($135\pm18$)~km~s$^{-1}$ which increases to ($165\pm43$)~km~s$^{-1}$\ only. In both the
cases, the period of oscillations was in between 15 min to 20 min. These propagations in inter-plume
regions were traced to originate from a bright region of the on-disk 
part of the coronal hole where the propagation speed is in the range of 
($25\pm1.3$)~km~s$^{-1}$\ to ($38\pm4.5$) km~s$^{-1}$ and increases to ($130\pm 51$)~km~s$^{-1}$ in off-limb, with
 the same periodicity, see Figure~\ref{fig:xt_ne8}. In this bright
region, the oscillations were seen in many spectral lines having a different formation temperature 
hence different height in the atmosphere. The observed oscillations in different lines have time
 delays with respect to the oscillations seen in lines forming lower in the atmosphere such as 
He~{\sc ii} 256~\AA\ and O~{\sc iv} 790~\AA\ which indicated the presence of upward propagation.
 From these \citet{2010ApJ...718...11G}
 have concluded that these waves were generated somewhere lower in the atmosphere,
probably in the chromosphere and then propagate upward and towards the off-limb region.  

A comparison between the two maps of both off-limb regions, indicate 
that the waves within the plumes were not observable (may be getting dissipated) 
far off-limb whereas that was not the case in the inter-plume region. \citet{2010ApJ...718...11G} have
interpreted these waves in inter-plume to be Alfv\'{e}nic based on the evidence that these obtained
propagation speeds were similar in lines formed at different temperature for some overlapping region
 and the propagation speed profile was found to be roughly consistent with that predicted for 
Alfv\'en waves which became supersonic ($>C_{s}\approx170$~km~s$^{-1}$, for Fe~{\sc xii} 195~\AA\
 line formation temperature) in far off-limb region. The presence of oscillations in line width
 with same range of periodicity as reported were also explained by these Alfv\'en waves. 
The radiance oscillations were 
explained due to the non-linear effects in density stratified atmosphere. Moreover, the measured propagation
 speeds were also consistent with the fast magnetoacoustic mode of propagation within the error bars
 of the propagation speeds in Ne~{\sc viii} 770~\AA\ and Fe~{\sc xii} 195~\AA\ lines and can explain the observed
 radiance oscillations due to its compressible nature. Hence, interpretation of these propagating
 disturbances in terms of fast magnetoacoustic waves were also indicated/supported.
Whereas the propagating waves in plume region were
interpreted as slow magneto-acoustic waves. The similar interpretation was also obtained by
\citet{2009A&A...499L..29B} for a plume region. \citet{2009A&A...499L..29B} have also studied the
x-t map obtained from Ne~{\sc viii} 770~\AA\ and Fe~{\sc xii} 195~\AA\ spectral lines simultaneously in plume region, 
(see Figure~\ref{fig:XTapr}) and obtained the propagation speed of 75~km~s$^{-1}$ and 125~km~s$^{-1}$ respectively
in both the spectral lines with period of oscillations between 10 min to 30 min. From these studies,
it can be conjectured that the inter-plumes are preferred channel for the acceleration of fast solar
wind due to the presence of Alfv\'{e}n waves which carries energy very far in corona without getting
dissipated whereas plumes have slow magneto-acoustic waves which are compressive in nature and can
 get dissipated by forming shocks in inner corona. 
Models of waves in plumes show that the waves are damped
over a distance of 0.25~R/R$_{\odot}$ above the photosphere \citep{2001ApJ...549L.143C}. In plumes
 in the lower corona, Coulomb collision are important, and
for temperatures of about 1 MK compressive viscosity is an effective dissipation
mechanism. Thermal conduction can also contribute significantly to the damping
of slow waves in coronal structures.

\citet{2006A&A...452.1059O} have studied the propagating waves in off-limb polar coronal holes
using CDS/SoHO in a statistical manner (described in section~\ref{sec:stats}).
In that study, they have measured the propagation speed between different line pairs, e.g.,
 O~{\sc v} 629~\AA\, Mg~{\sc x} 609~\AA\ \& 625~\AA\ and Si~{\sc xii} 520~\AA\. The measured speed between O~{\sc v} and Mg~{\sc x}
 lines was ($154\pm18$)~km~s$^{-1}$ whereas that between O~{\sc v} and Si~{\sc xii} lines was
($218\pm 28$)~km~s$^{-1}$. These measured propagation speeds were close to the sound speed
for the temperature of formation of these ions, suggesting the presence of slow magneto-acoustic 
waves. The measured oscillations in their analysis were present over the
 frequency range of $\approx$~0 mHz to 8 mHz. They have also reported the evidence of fast magneto-acoustic
waves at coronal temperatures based on the histogram of flux-velocity (I--V) phase delay plot. They
further conjectured that these waves may have been generated in the upper atmosphere in and around
the coronal temperatures at which the Mg~{\sc x} and Si~{\sc xii} lines form. 

Up to now evidence for the fast
magnetoacoustic wave modes in these coronal hole regions has been very limited,
even though recent results by \citet{2005A&A...430L..65V} have shown that
propagating fast magnetoacoustic waves can be present in open
magnetic field structures, albeit in this instance, in a post-flare
supra-arcade. For the fast mode the wavelengths of the propagating
wave should be much shorter than the size of the structure guiding
the wave. Shorter wavelength implies shorter period, thus it
demands high cadence observations. TRACE can work at 20 s to 30 s
cadence, allowing us to detect waves with a 40 s to 60 s periodicity at
the best. Thus, it is difficult to detect smaller periodicity with the
present space-based instruments, whereas ground-based coronagraphs
and radioheliographs have much better time resolution and have been
used for the detection of short-period waves. The Solar Dynamic Observatory
 should allow us to detect short periodicity with its high cadence imaging instrument AIA.

\subsection{Detection of Alfv\'{e}n Waves from Variation of Line Width}
As already mentioned, wave-induced motions on unresolved scales result in a broadening of the
 observed spectral profiles. The measured
broadening of the optically thin spectral lines of ions is due to two
effects, thermal broadening and non-thermal broadening associated
with Doppler shifts due to the unresolved line-of-sight motions
\begin{equation}
 T_{\rm eff}=T_i+\alpha\frac{\rm m_i}{2k}<v_{\rm LOS}^2> 
\end{equation}

where $T_i$ is the ion temperature, $k$ is the Boltzmann constant,
$v_{\rm LOS}$ is the line-of-sight component of the velocity, and,
$2/3\le\alpha\le 1$. There have been several off-limb spectral line observations to 
search for Alfv\'{e}n wave signatures. \citet{1977ApJ...212L.143D} have reported the 
presence of non-thermal velocities of about $\approx$20~km~s$^{-1}$ above the limb 
($\approx30\arcsec$) in coronal hole region in Si~{\sc viii} 1446~\AA\ spectral line recorded by the NRL
 slit spectrograph on Skylab. Measurements of ultraviolet 
Mg~{\sc x} 609\AA\ \& 625~\AA\ line widths during a rocket flight showed an increase in 
width with height to a distance of 70\,000~km \citep{1990ApJ...348L..77H}. Also, coronal hole 
Fe~{\sc x} 6374~\AA\ spectra taken at Sacramento Peak Observatory    
showed an increase in the line width with height \citep{1994SSRv...70..373H}. 
SUMER/SoHO were used to record the off-limb, height-resolved spectra of an Si~{\sc viii}
(1446~\AA\ and 1440~\AA) density sensitive line pair, in  equatorial coronal regions 
\citep{1998SoPh..181...91D,2005A&A...435..733W} and in polar coronal holes 
\citep{1998A&A...339..208B,2004A&A...415.1133W}. \citet{1997SoPh..173..243D} and \citet{1998A&A...337..287E} have also
reported the presence of nonthermal velocities in chromospheric, transition region and coronal lines
for equatorial region.

\begin{figure}
\centering
\includegraphics[width=11cm]{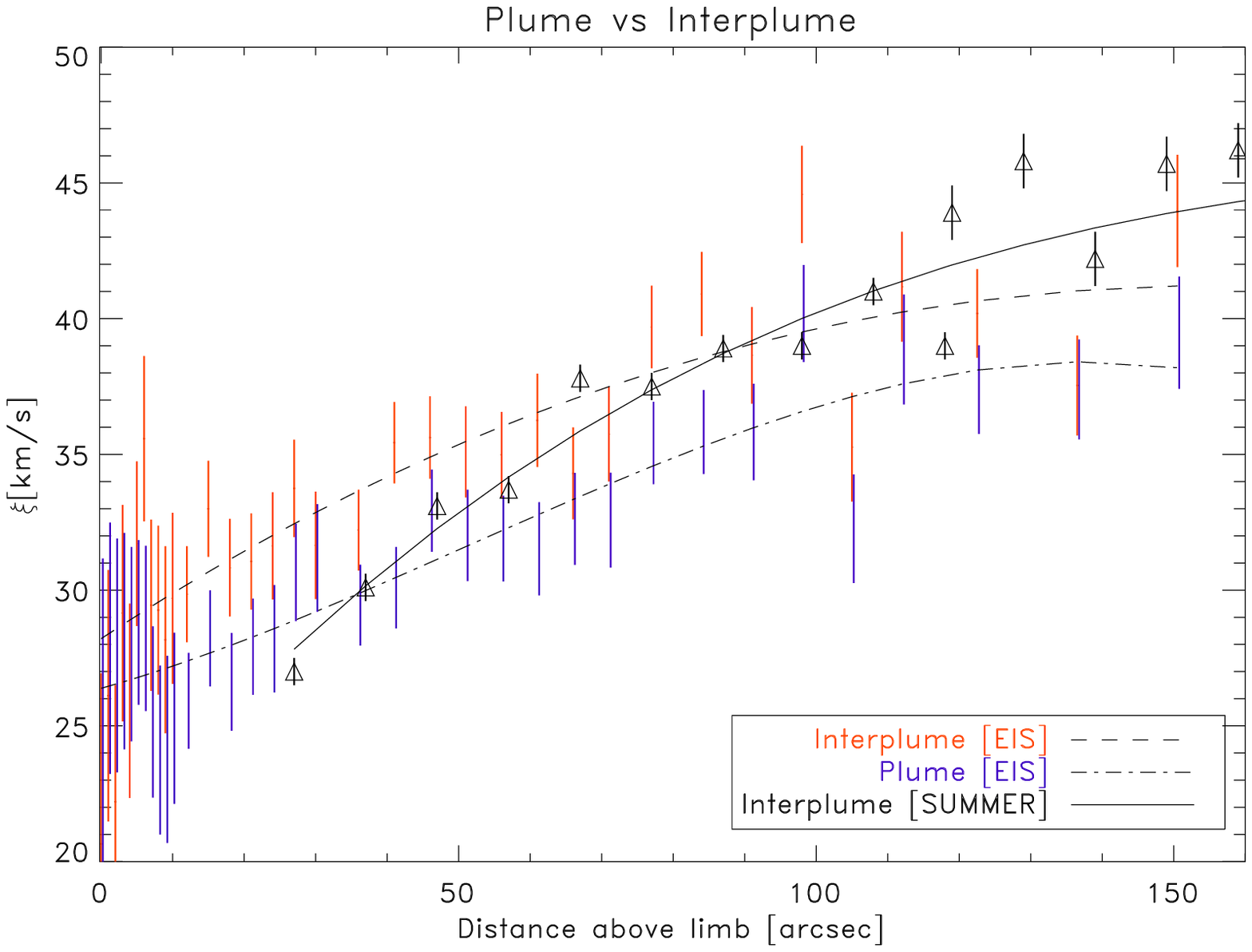}
\caption {Variation in nonthermal velocity with height as recorded by 
Fe~{\sc xii} 195~\AA\ along a polar plume and interplume. 
The solid line corresponds to the nonthermal velocity as derived from  
Si~{\sc viii} 1446~\AA\ \citep{1998A&A...339..208B}. The dashed 
line is a third-order polynomial fit. From Banerjee et al. (2009b)} \label{fig:comparesumer}
\end{figure}
The measured variation in line width with 
density and height supports undamped wave propagation in coronal structures. This 
was strong evidence of outwardly propagating 
undamped Alfv\'{e}n waves in the corona, which may contribute to coronal 
heating and high-speed solar wind in the case of coronal holes. 
\begin{figure}
\centering
\includegraphics[width=11cm]{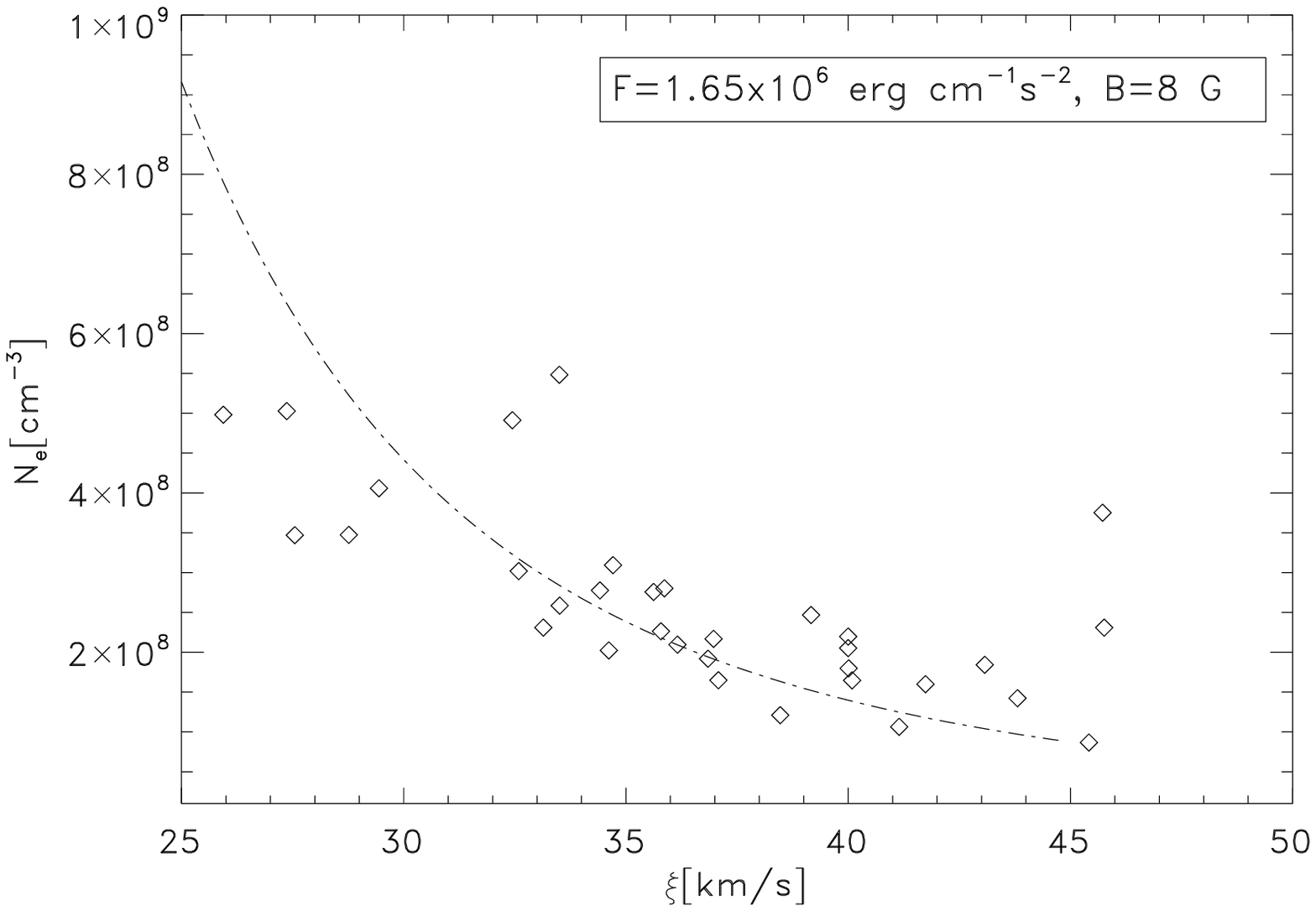}
\caption {Variation in electron density with nonthermal velocity for the
polar coronal hole. The squared boxes represents 
the observed values and the solid line represents the theoretical relation 
(Eq.~[4]) for fixed magnetic field strength as indicated. From Banerjee et al. (2009b)} \label{fig:relation}
\end{figure}
Alfv\'en waves propagate virtually undamped through the quasi-static corona and 
deposit their energy flux in the higher corona. The total energy flux crossing 
a surface of area $A$ in the corona due to Alfv\'en waves is given by 
\citep{2001A&A...374L...9M}
\begin{equation}
F = \sqrt{\frac{\rho}{4 \pi}} <\delta v^2> B\ A ~~, 
\end{equation}
where $\rho$ is the plasma mass density (related to $N_{\rm e}$ as $\rho\approx m_{\rm p} N_{\rm e}$,
$m_{\rm p}$ is the proton-mass), $<\delta v^2>$ is the mean square velocity related 
to the observed nonthermal velocity as $ \xi^2 = <\delta v^2>/2 $, and 
$B$, is the magnetic field strength. If the wave energy flux is conserved as 
the wave propagate outwards, then Eq.(2) gives
\begin{equation}
<\delta v^2>^{1/2} \ \propto \ \rho^{-1/4} (BA)^{-1/2}~~,
\end{equation}
Now if one assumes a flux tube geometry, then $BA$ is constant with height and 
Eq.(3) yields 
\begin{equation}
<\delta v^2>^{1/2} \ \propto \ \rho^{-1/4}~~.
\end{equation}
To explore the validity of Eq.(4), as done by \citet{1998A&A...339..208B}, 
 \citet{2009A&A...501L..15B} 
 plotted Figure~\ref{fig:relation} corresponding to interplume data. The solid lines are the 
theoretically predicted functional forms of 
the variation of number density with nonthermal velocity (Eq.[4]) and the 
diamonds are the observed data. The proportionality constant have been chosen 
to match the calculated energy flux. For the interplume data they have used $B$=8 G at certain height. 
Once again the agreement is very good, especially when we are away from 
the limb. Thus it becomes possible to detect Alfv\'en waves from the study of the variation of line widths.
\citet{2008A&A...483..271D} have also reported detecting the Alfv\'en waves from the study of the variation of
 line widths in solar off-limb region, where they do not find evidence for damping of
the Alfv\'en waves. It will be here interesting to point out that \citet{1997SoPh..173..243D} have
also reported the presence of nonthermal velocities in quiet Sun and coronal hole regions using the
fourth rocket flight of High Resolution Telescope and Spectrograph (HRTS-4) instrument. We
 should also point out that the additional line widths are generaly interpreted in terms of
 wave-induced motions on unresolved scales but there are two other possibilities; (i) increasing
 ion temperatures with off-limb distance \citep{1998ApJ...503..475T}, or additional flow components
 within the FOV due to spicules/macro-spicules flows \citep{2005A&A...431L..17D}. Are there Alfv\'en waves present in the lower solar atmosphere? This has been questioned by \citet{2007Sci...318.1572E} while looking at the initial results from Hinode. \citet{2009Sci...323.1582J}  reported the detection of oscillatory phenomena
of line widths (FWHM) associated with a large bright point group located near the solar disk center which indicates
the presence of Alfv\'en waves in that region. These studies have focussed on the detection of the oscillatory nature of the line widths at fixed location rather than variation with height.

\section{Propagating Waves in on-disk region of coronal holes}
\label{sec:on-disk}

\begin{figure}
\centering
\includegraphics[width=12cm]{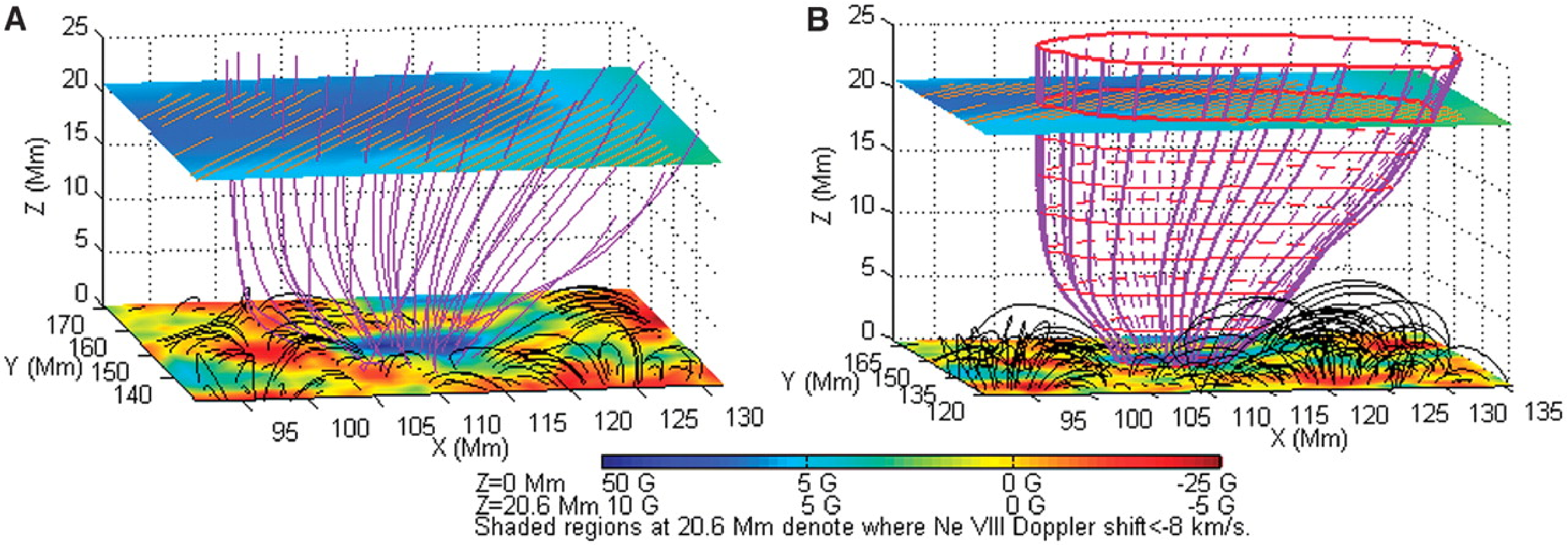}
\caption {Magnetic funnel in the solar atmosphere. (A) Emphasis on open field lines and correlation
 with the Ne~{\sc viii} 770~\AA\ outflow speed larger than 8~km~s$^{-1}$ (dark shading). (B) Illustration of the
 funnel boundary and magnetic unipolar flux constriction by adjacent, surrounding bipolar
 loops. From \citet{2005Sci...308..519T} } \label{fig:Tu}
\end{figure}

In the previous sections we have discussed the observational evidences of the presence of
 propagating MHD waves. Now we would look for their origin. These waves must be generated
 somewhere within the disk part of the coronal hole. 
Similar to the quiet Sun, a network and internetwork pattern can also
be seen all over the coronal holes in chromospheric and transition region lines.
 The magnetic field is predominantly concentrated on these network regions and the
footpoints of coronal funnels emanate from these regions, see Figure~\ref{fig:Tu} 
 \citep{2005Sci...308..519T}. All the
 open field lines forming the coronal holes essentially originate from such regions and the fast
 solar winds emanate from these vertical flux tubes. \citet{2008ApJ...688.1374T} have used the
high spatial resolution of Solar Optical Telescope aboard Hinode to characterize the magnetic 
landscape of the Sun's polar region. They have conjectured that the vertical flux tubes originating
from the kilogauss magnetic patches in coronal holes will fan out in the lower atmosphere of the 
coronal holes. All these patches have the same polarity which is consistent with the global polarity
of the polar region. They have further conjectured that these vertical flux tubes with large expansion
around the photospheric-coronal boundary (which may be the upper chromosphere) will serve as
 efficient chimneys for the propagation of 
Alfv\'{e}n waves which accelerate the solar wind. The location of these magnetic patches is still
unclear but most likely, are placed on network boundaries within the coronal holes. The nature of
waves in these regions are being studied now and some recent results are discussed in the next section. 
\subsection{Detection of Waves using statistical techniques}
\label{sec:stats}

\begin{figure*}[htbp]
 \centering
\includegraphics[angle=90, width=10cm]{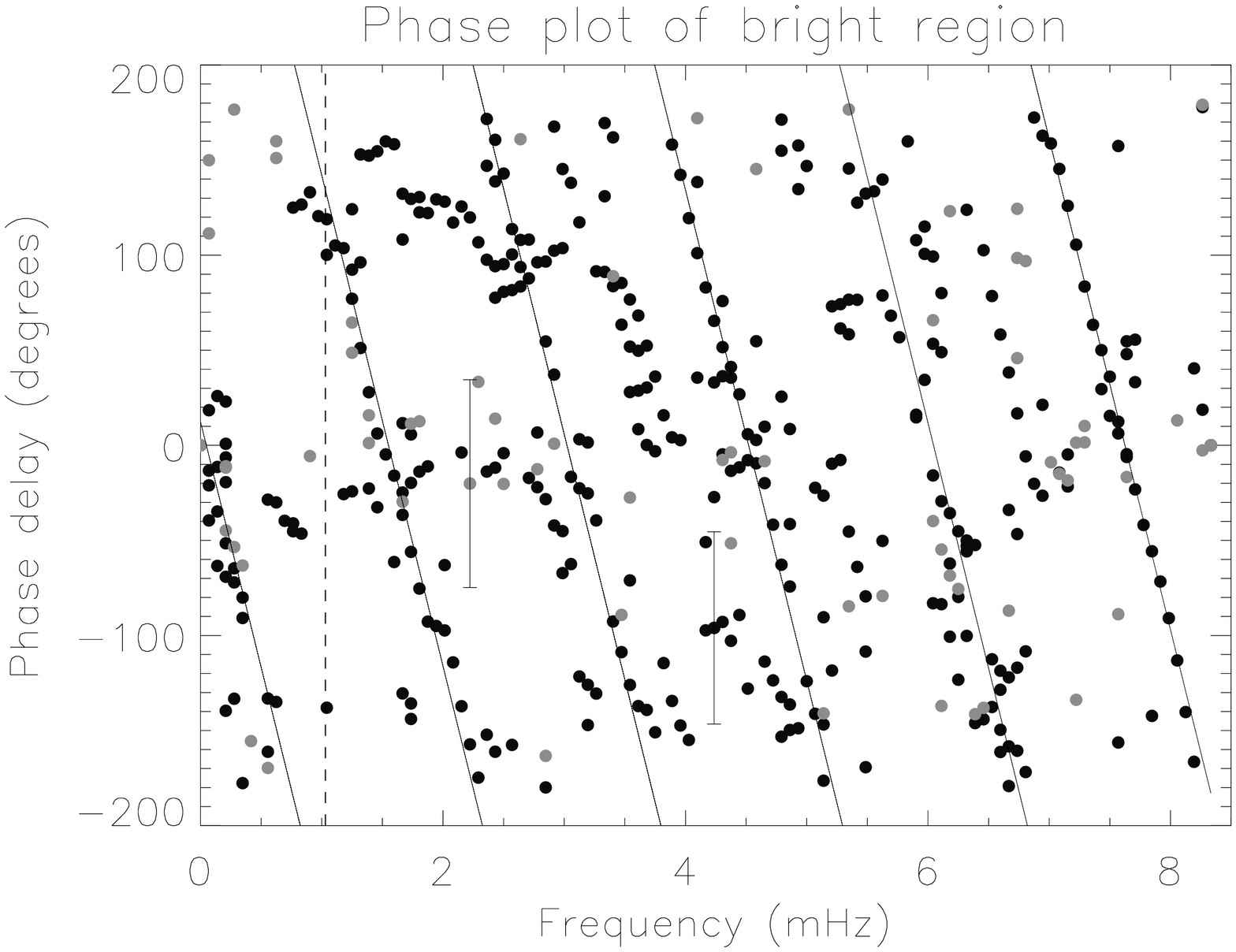}\\{\includegraphics[angle=90, width=10cm]{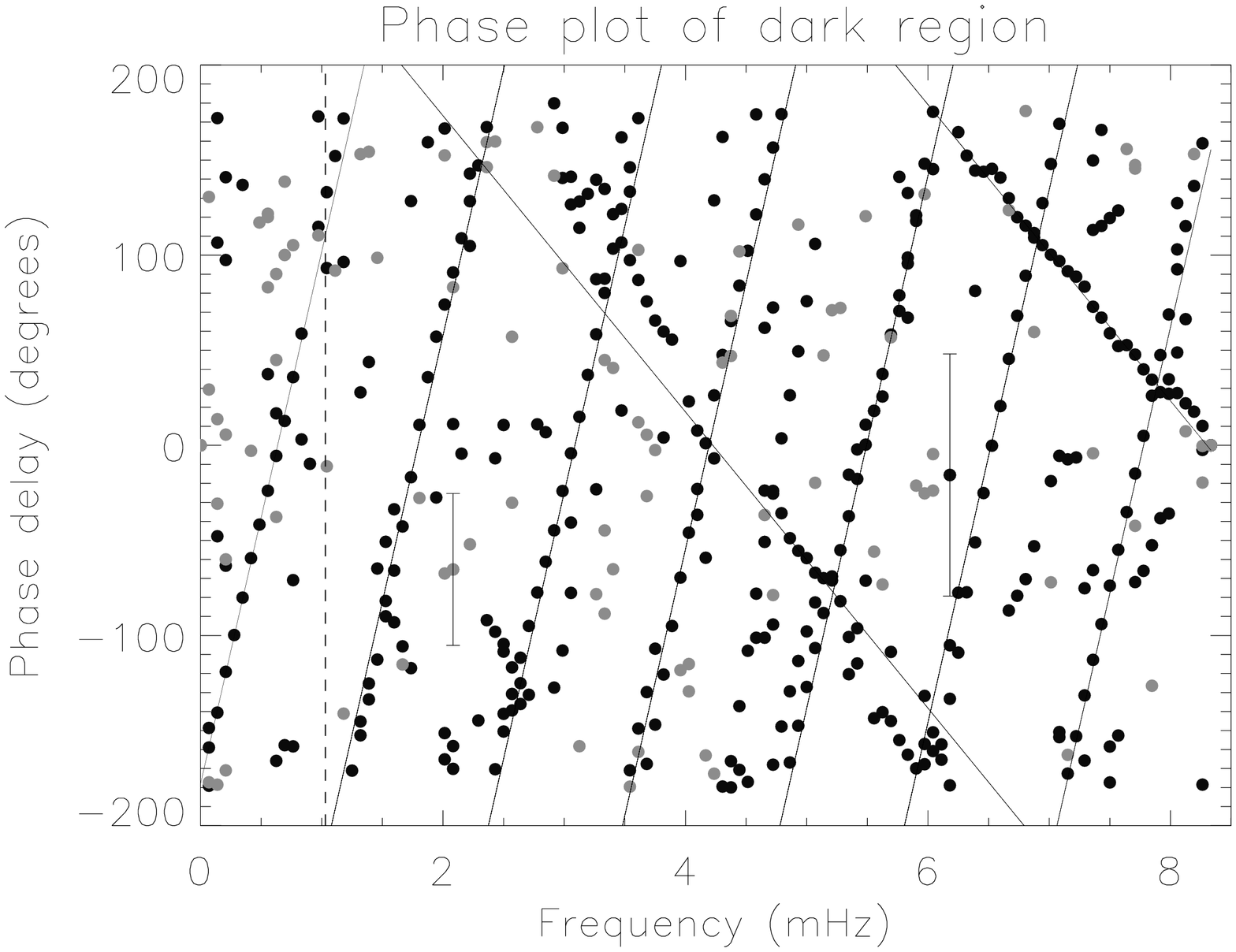}}
 \caption{Phase delays measured between the oscillations in the
 spectroscopic line pair for the bright (top panel) and dark (bottom panel) locations. Presumably,
 the bright locations correspond to the network boundaries. The phases
 in radiant flux oscillations are shown in the grey circle symbols
 while that in LOS velocities are shown as the black circle
 symbols. Overplotted on each Figure are black  parallel lines,
 corresponding to fixed time delays. The vertical dashed line drawn at 1.03 mHz indicates that
 some phase values below this could be
affected by solar rotation. Representative errors on the phase measurements are indicated by
 the error bars. From \citet{2009A&A...493..251G}.}
 \label{fig:phase}
\end{figure*}

\citet{2009A&A...493..251G} had studied the nature of propagating waves in these
network and internetwork regions using SUMER polar coronal hole  data in a
statistical manner. They have divided the dataset in those two regions based on the intensity
 enhancements. They have seen the similar kind of oscillations in two spectral lines of N~{\sc iv}
765~\AA\ and Ne~{\sc viii} 770~\AA\ in intensity as well as in velocity. Phase delays were measured using the 
technique of \citet{1979ApJ...229.1147A} and plotted the phase delays over the full range of 
$-180\degree$ to $180\degree$ as a function of frequency. An example of the result of this is shown
in Figure~\ref{fig:phase} from \citet{2009A&A...493..251G}.

In this Figure, the phase delays measured between the line pair for both the network and internetwork
regions are shown. From Figure~\ref{fig:phase}, it can be seen that the measured oscillations are 
present over the frequency range of $\approx~0$ mHz to 8 mHz. It can be also seen that these phases line up
along parallel straight lines. This distribution of phases along straight lines indicates the 
presence of upward or downward propagating waves depending upon the inclination of straight line.
From the measurement of slope of these lines one can obtain the time delays between the different
 lines, using the phase equation,
\begin{equation}
 \Delta{\rm \phi} = 2\,\pi\,\textit{f}\,T
\end{equation}
where $\textit{f}$ is the frequency and T the time delay in seconds,
the phase difference will vary linearly with \textit{f}, and will
change by $360$ over frequency intervals of $\Delta \textit{f} =
1/T$. This will give rise to parallel lines in $\Delta{\rm \phi}$
vs. $\textit{f}$ plots at fixed frequency intervals ($\Delta
\textit{f} =1/T$), corresponding to a fixed time delay T. 
In the case of Figure~\ref{fig:phase}, The phase plot for network region (left panel) shows only the
presence of upwardly propagating waves whereas that for internetwork region (right panel) shows
the presence of both upward and downward propagating waves. The time delay measured for the two
spectral lines was found to be ($717\pm 114$)~s for the network regions whereas that for the
internetwork regions were ($216\pm 38$)~s and ($778\pm 133$)~s for the upwardly and downwardly
propagating waves respectively. Using the limb-brightening technique, the height difference between
the spectral lines were measured and using these time delays, \citet{2009A&A...493..251G} have
calculated the upwardly propagation speed of ($4.0\pm 0.6$)~km~s$^{-1}$ in network region whereas
upwardly and downwardly propagation speeds were  ($13.2\pm 2.3$)~km~s$^{-1}$ and 
($3.7\pm 0.5$)~km~s$^{-1}$ respectively in internetwork region. These measured speeds suggest the presence of
slow magneto-acoustic waves in these on-disk regions of coronal hole. \citet{2009A&A...493..251G}
 have conjectured that the internetwork regions within coronal holes are composed of
low-lying coronal loops, where waves can travel both upwardly and
downwardly, whereas the network regions are filled with more open,
funnel like structures where only upward propagation is possible (see Figure~\ref{fig:Tu}).
 It has also been point out
that statistically there were more velocity oscillations than intensity
oscillations in the phase plots. For compressional types of waves,
 signatures are seen in both. So if there are only velocity
oscillations present in some locations this implies the presence of transverse-type waves,
 like Alfv\'{e}nic. It has been further pointed out that in this particular study the identified
 network region spans
a large spatial domain along the slit, which would be due to the bigger magnetic patch. 
\begin{figure}[htbp]
\centering
\includegraphics[angle=90, width=12cm]{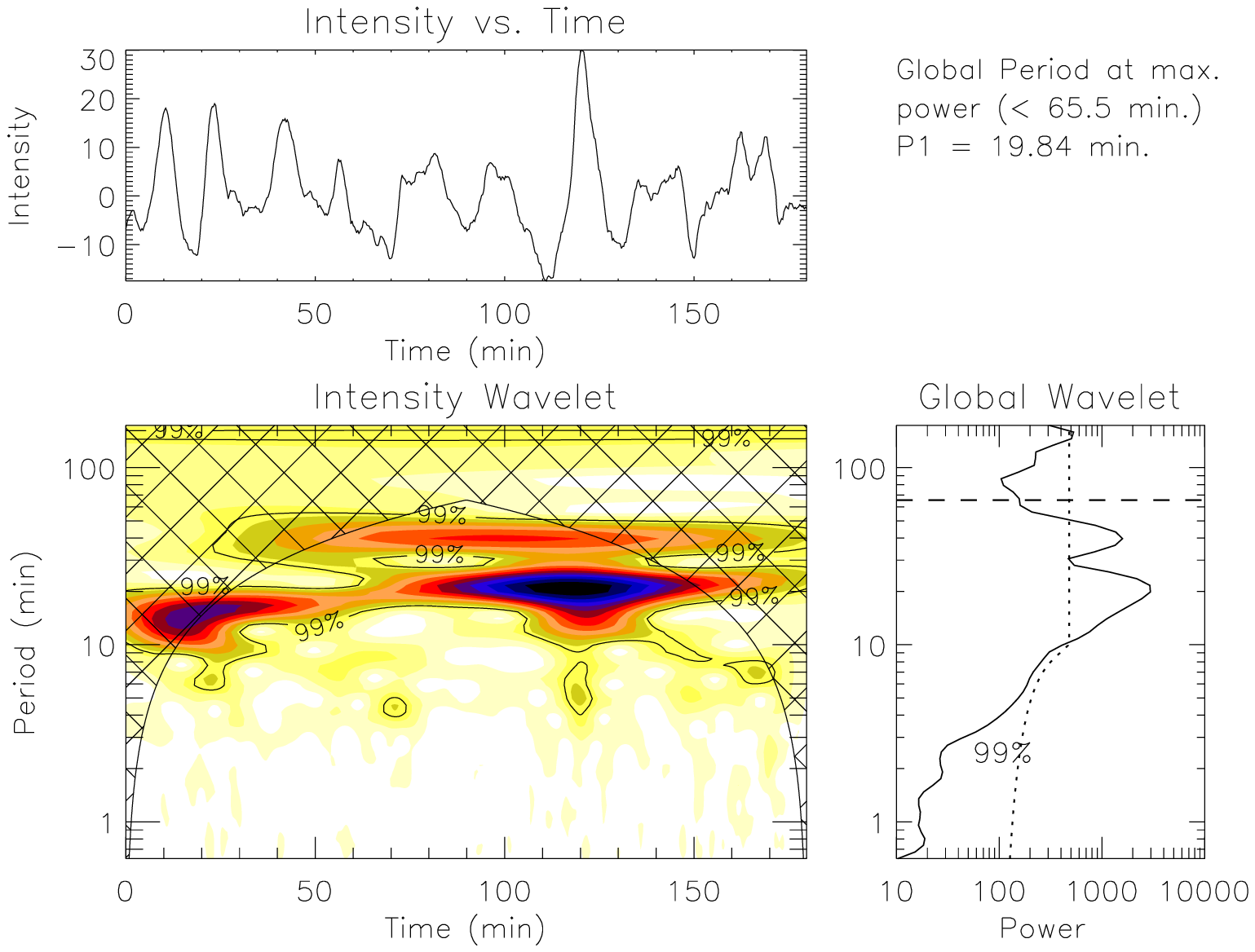}
\caption{The wavelet result for the on-disk location at
 solar-Y$\approx 967\arcsec$ and solar-X$\approx -72\arcsec$ in Ne~{\sc viii} radiance
 (left side) and velocity (right side). The
 top panel shows the relative (background trend removed) radiance 
 smoothed over 3~min. Bottom left panel shows the color inverted wavelet power
 spectrum with $99~\%$ confidence level contours while
 bottom right panel show the global (averaged over time) wavelet power spectrum with $99~\%$
 global confidence level drawn. 
 The period P1 at the location of the maximum in the global wavelet spectrum is 
 printed above the global wavelet spectrum. 
From \citet{2010ApJ...718...11G}. \label{fig:wavelet}}
\end{figure}
As found in \citet{2010ApJ...718...11G} (see also section~\ref{sec:xtmap}), the waves propagating
 in the off-limb region
which has its origin in on-disk bright region can also be linked to the magnetic patch on
network boundary. The nature of oscillation observed on that region is very much similar to that
observed by \citet{2009A&A...493..251G} in this case. Figure~\ref{fig:wavelet} shows the kind of
oscillations observed by Gupta et al. (2010) in the bright region in the Ne~{\sc viii} 770~\AA\ spectral
line radiance. The similar oscillations are also present in Ne~{\sc viii} 770~\AA\ Doppler velocity. The
generation mechanism of these oscillations is still unclear. Overall these regions are very much
similar to the magnetic patches reported by \citet{2008ApJ...688.1374T} and hence will serve as the
efficient chimneys for Alfv\'{e}n waves which will accelerate the solar wind.
\begin{figure}[!th]
\centering
{\includegraphics[width=8cm, clip=true]{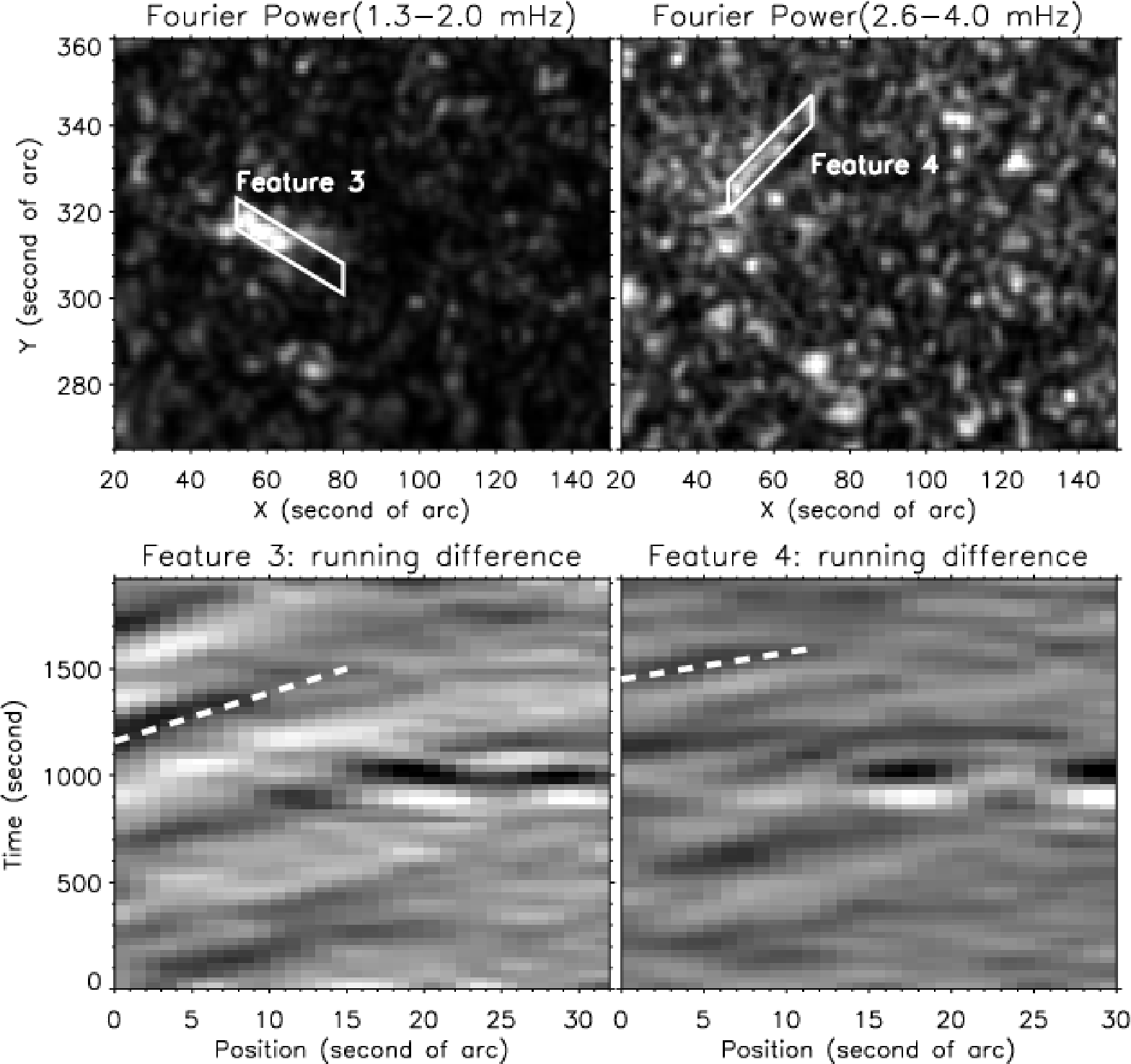}}\\
\caption{Upper: Maps of Fourier power in frequency ranges of 1.3 mHz to 2.0 mHz and 2.6 mHz to
 4.0 mHz for the 171~\AA\ passband. Lower: plots of running differences along the long sides of
 the bars shown in the upper panels. The dashed lines used to calculate the speed of propagating
 oscillations. From \citet{2008A&A...488..331T}.}
\label{fig:xt_net}
\end{figure}
\subsection{Waves in equatorial coronal holes}
\label{sec:equat}
So far we were discussing on polar coronal holes. Coronal holes are often found, particularly during
 solar maximum in the equatorial region also. \citet{2008A&A...488..331T} have studied the network
 oscillations in equatorial coronal hole. For the
analysis, two bandpass (1600~\AA and 171~\AA) time series images were obtained from the TRACE 
spacecraft. They have measured the oscillations of period in between 5 min to 10 min in network 
enhanced regions from wavelet analysis. Using the time-distance map for the same region, the
 propagation speeds were measured to be 32~km~s$^{-1}$ for the 10 min oscillation and 58~km~s$^{-1}$
for the 5 min oscillation, see Figure~\ref{fig:xt_net}. \citet{2008A&A...488..331T} have interpreted
 these propagations to be slow magneto-acoustic waves and calculated the energy
flux of 40 erg~cm$^{-2}$~s$^{-1}$ for the quiet coronal conditions which are much lower than the
 required heat input to the quiet corona. Whereas for the chromospheric conditions, the calculated
energy flux is 1.368$\times$10 $^{6}$ erg~cm$^{-2}$~s$^{-1}$ which is almost one third of the 
required energy budget for the chromosphere.\\ 
\citet{2007A&A...463..713O} have also studied the propagating waves in equatorial as well as in
polar coronal holes using CDS/SoHO. They have also analyzed these
 propagating waves in a statistical manner which is explained in section~\ref{sec:stats}. In their
study, they have measured the propagation speed between different line pairs, e.g., O~{\sc v} 629~\AA\, 
Mg~{\sc x} 624~\AA\ and Si~{\sc xii} 520~\AA. The obtained propagation speed between these line pairs was in between
50~km~s$^{-1}$ to 120~km~s$^{-1}$ and the measured oscillations were present over the frequency range of 
$\approx$~0 mHz to 8 mHz. They have estimated the energy carried by these waves in WKB approximation, and 
obtained the wave energy flux of $\approx 2.5\times10^{4}$  erg~cm$^{-2}$~s$^{-1}$. This is lower 
than the energy flux requirement of a coronal hole with the high speed wind ($\approx 8\times10^{5}$
  erg~cm$^{-2}$~s$^{-1}$), but non-WKB effects may enhance the energy flux somewhat. 
\citet{2007A&A...463..713O} have also found that these waves were propagating preferentially in 
regions of increased brightness in the coronal holes that are considered to be locations of 
concentrated magnetic field which could be the high field magnetic patches as reported by 
\citet{2008ApJ...688.1374T}.

\subsection{The source of the propagating waves}
In order to answer the question of where propagating
coronal waves originate from, \citet{2004Natur.430..536D,2006RSPTA.364..383D} developed the
 general framework of how photospheric oscillations can leak into the atmosphere along
inclined magnetic-flux tubes. In a non-magnetic atmosphere $p$\,modes
are evanescent and cannot propagate upwards through the temperature
minimum barrier since their period $P$ ($\approx$~200 s to 450 s) is above
the local acoustic cut-off period $P_c \approx 200$ s. However, in a
magnetically structured atmosphere, where the field lines have some
natural inclination ($\theta$), where $\theta$ is measured between the
magnetic guide channelling the oscillations and the vertical, the
acoustic cut-off period takes the form $P_c \sim
\sqrt{T}/\cos{\theta}$ with the temperature $T$. This inclination
will allow some non-propagating evanescent wave energy to tunnel
through the temperature minimum into the hot chromosphere of the
waveguide, where propagation is once again possible because of
higher temperatures ($P_c > 300$ s). There has been recent attempts  to numerically model the direct propagation of acoustic waves, driven harmonically at the solar photosphere, into the three-dimensional solar atmosphere  in the framework
of ideal magnetohydrodynamics \citep{2007A&A...467.1299E,2007SoPh..246...41M,2009SoPh..258..219F}. They have studied the leakage of
5 min global solar acoustic oscillations into the upper, gravitationally stratified and magnetised
atmosphere, where the modelled solar atmosphere possesses realistic temperature
and density stratification. They have shown that high-frequency waves can propagate from the
lower atmosphere across the transition region, experiencing relatively low reflection, and
transmitting most of their energy into the corona. The thin transition region acts as a
wave guide for horizontally propagating surface waves for a wide range of driver periods. One must point out that this modelling has been restricted to either for simple extended flux tubes or non magnetic atmosphere. 
More realistic models are needed to verify some of these calculations.

\section{Concluding remarks}
There is direct evidence of the presence of  magnetoacoustic waves in
solar coronal holes obtained with EUV imagers and spectroscopy (time-distance radiance maps). The wave
relative amplitudes are several percent and periods are several minutes. Evidence for Alfv\'{e}n and fast magnetoacoustic waves is indirect but abundant in both imaging and spectral data. The source of these waves and the
physical mechanism responsible for the observed periodicity has not yet been fully understood. Theoretical progress in this field is very much needed to explain the observed signatures. 

Recently \citet{2010A&A...510L...2M} have analyzed STEREO observations of plumes and interpreted them in terms of  quasi-periodically forced jets. In almost all polar plumes observed at the limb in these STEREO sequences, in all coronal passbands, they observe radiance variations travelling along the structures with a mean velocity of 135~km~s$^{-1}$
 at a range of temperatures from 0.5~MK to 1.5~MK, that they interpret as high speed jets of plasma. The jets have an apparent brightness enhancement of 5~\% above that of the plumes they travel on and repeat quasi-periodically, with repeat-times ranging from five to twenty-five minutes. They also claim that these
perturbations have very similar properties to those observed in the quiet Sun by \citet{2009ApJ...707..524M}. In each
case, these perturbations have been connected spectroscopically to a strong upflowing, weak emission component at the magnetic
footpoints. The spectroscopically determined upflows appear to be rooted in dynamic \textit{Type-II} spicules in the upper chromosphere
\citep{2009ApJ...701L...1D}. It should be pointed out here that this alternative interpretation is
 contrary to the widely held interpretation that this observational phenomenon is due to compressive
 waves, nonetheless it should not be ruled out completely but a further analysis and new set of 
observation in equatorial coronal hole is desired (as indicated by \citet{2010A&A...510L...2M})
 to  address the wave/upflow interpretation issue. Often the non-detection of radiance variations at larger heights are due to poor signal to noise, so future imaging instruments with better detectors can also address this issue. One should also consider Doppler dimming results
 from UVCS/SoHO and SUMER/SoHO, which shows that velocities above 100~km~s$^{-1}$ are reached only above
 1.5~R/R$_{\odot}$ in either plumes or interplumes \citep{2003ApJ...588..566T,2003ApJ...589..623G,2005ApJ...635L.185G}.
 Besides, there is no evidence of such large flows in accurate Doppler maps of
 the off-limb region of coronal holes \citep{1998ApJ...500.1023W,teriaca2010}.

The future of wave detection looks very promising . With the  Hinode satellite, containing the X-Ray Telescope (XRT)
and the EUV Imaging Spectrometer (EIS), and the just launched Solar Dynamics Observatory (SDO), containing the
Atmospheric Imaging Assembly (AIA),  means that it will soon be
possible to obtain slit and image data at a much increased spectral
resolution with excellent time resolution. For example, the
AIA, offering a replacement for the TRACE instrument, will allow the Sun to be imaged at ten different
wavelengths, eight of them simultaneously, with a time resolution of $\approx$~10~s.
 The good spectral resolution of EIS
will, in addition, allow the accurate measurement of non-thermal
velocities and allow other studies that are based on the detailed
measurement of line widths,  together with a simultaneous
measurement of electron density in these structures.

\begin{acknowledgements}
We like to thank Prof.  Klaus Wilhelm for careful reading of the manuscript.
\end{acknowledgements}

\bibliographystyle{aps-nameyear} 
\bibliography{references.bib}   

\end{document}